\documentclass[USenglish]{article}

\usepackage[utf8]{inputenc}
\usepackage[small]{dgruyter}
\usepackage{amssymb}
\usepackage{amsmath}
\usepackage{exscale}
\usepackage{mathrsfs}
\usepackage{array}
\usepackage{graphicx}
\usepackage{caption}
\usepackage{subcaption}

\begin{document}
  \author[1]{Wolfgang Karl H{\"a}rdle}
  \author*[2]{Elena Silyakova}
  \affil[1]{Ladislaus von Bortkiewicz Chair of Statistics, Humboldt-Universit{\"a}t zu Berlin, Unter den Linden 6, 10099 Berlin, Germany, Sim Kee Boon Institute for Financial Economics, Singapore Management University Administration Building, 81 Victoria Street, 188065 Singapore. Email: haerdle@hu-berlin.de.}
  \affil[2]{Ladislaus von Bortkiewicz Chair of Statistics, Humboldt-Universit{\"a}t zu Berlin, Unter den Linden 6, 10099 Berlin, Germany. Email: silyakova@gmail.com}
  \title{Implied Basket Correlation Dynamics}
  \runningtitle{Implied Basket Correlation Dynamics }
  \abstract{Equity basket correlation can be estimated both using the physical measure from stock prices, and also using the risk neutral measure from option prices. The difference between the two estimates motivates a so-called ``dispersion strategy''. We study the performance of this strategy on the German market and propose several profitability improvement schemes based on implied correlation (IC) forecasts. Modelling IC conceals several challenges. Firstly the number of correlation coefficients would grow with the size of the basket. Secondly, IC is not constant over maturities and strikes. Finally, IC changes over time. We reduce the dimensionality of the problem by assuming equicorrelation. The IC surface (ICS) is then approximated from the implied volatilities of stocks and the implied volatility of the basket. To analyze the dynamics of the ICS we employ a dynamic semiparametric factor model.}
  \keywords{correlation risk, dimension reduction, dispersion strategy, dynamic factor models}

  \startpage{1}
  \aop

\maketitle

\section{Introduction}\label{sec:intro}
Equity basket correlation is an important risk factor. It characterizes the strength of linear dependence between assets and thus measures the degree of portfolio diversification. It is an input for many pricing models, plays a key role in portfolio optimization and risk management. The concept of a time-varying correlation is frequently used in studies that describe the joint dynamics of assets, \cite{Bollerslev1998}, \cite{Engle2002}. However, the idea of considering the correlation as an asset, on its own, is relatively new and has recently gained popularity together with the emergence of such derivative instruments as variance, volatility, correlation swaps and trading strategies with them, \cite{Demeterfi1999}, \cite{Bossu2005}. In this context being able to predict correlation patterns might help to reveal profitable trading opportunities. One of the most common ways of obtaining a correlation exposure is to replicate it with variance swaps. In this paper we study the behaviour of a particular vehicle for trading correlation known as a ``dispersion strategy'', in which one sells a stock index volatility and buys individual volatilities, \cite{Allen2005}. We propose several ways of improving the profitability of the strategy by extracting information from a dynamic model of implied correlation.

Unlike asset prices, correlations are not directly observed in the market and need to be estimated in the context of a particular model. Obtaining a well-conditioned and invertible  estimate of an empirical correlation matrix is often a complicated task, in particular when the dimensionality of basket elements $N$ is higher than the time series length $T$.  Here some work has been done in the field of random matrix theory (RMT), in which the case ``large $N$, small $T$'' is studied in an asymptotic setting, \cite{Bai1999}, \cite{LaurentLaloux1999}, \cite{VasilikiPlerou2002}.  A further segment of research has moved in the direction of developing various regularization methods for sample covariance and correlation matrices, such as a shrinkage technique proposed in \cite{Ledoit2003}, regularization via thresholding in \cite{Bickel2008a}, bending in \cite{Bickel2008}, factor models in \cite{Fan2008} and many others. There are some studies that propose a dynamic model for returns correlation such as a DCC model by \cite{Engle2002}, and in a high-dimensional setting, \cite{Engle2008}. The common feature of all these studies is that the empirical correlation matrix is estimated under the physical measure from the time series of asset returns. Alternatively, instead of relying on historical data, one can infer correlation from the current snapshot of the option market. Option prices reflect the expectations of market participants about the future price (volatility) and disclose their perceptions of market risk, \cite{Bakshi2000}, \cite{Britten-Jones2000}. Some recent studies have shown that the implied volatility (IV), that equates the model option price and the one taken from the market, contains incremental information beyond the historical estimate and outperforms it in forecasting future volatility, \cite{Christensen1998125},  \cite{Fleming1998317}, \cite{Blair20015}. Yet only a few papers have studied the predictive content of the correlation, implied by option prices.  Some work has been done for foreign exchange (FX) options, \cite{Campa1998855}, \cite{Lopez1998}, which showed that correlations implied from FX options are useful for forecasting future currency correlations. \cite{Skintzi2005} investigated the average correlation implied by equity options and introduced the Implied Correlation index (ICX). They showed that ICX, computed from current option prices, is a useful proxy for the future realized correlation. \cite{Vilkov2009} investigate the power of options implied correlation to explain the future realized correlation and conclude that its predictive power is quite high.

Here we model the implied correlation (IC), which is an object of very high dimensionality. Similarly to the IV, every day one recovers an IC surface. We model the IC with a dynamic semiparametric factor model (DSFM), \cite{Fengler2007}, \cite{Park2009} and \cite{Song2014}, and find that it yields a low dimensional representation as a linear combination of a small number of time-invariant basis functions (surfaces), whose time evolution is driven by a series of coefficients; technical aspects are also described in \cite{Sperlich1999}. We produce an IC forecast and use it in several hedging schemes for a dispersion strategy. For the empirical analysis we chose the German market represented by the DAX portfolio over the 2-years sample period from 20100802 to 20120801 (dates are written as YYYYMMDD). Backtesting shows that the hedge allows to the reduction of potential losses and increases the average profitability of the strategy.

The paper is structured as follows. In Section \ref{sec:corr} we introduce the notions of realized, model-implied and model-free implied volatility and correlation and describe the basic setup of a dispersion strategy with variance swaps. The DSFM model for IC is introduced in Section \ref{sec:cormod} starting with general description in Section \ref{ssec:model}, followed by the description of the functional principal component analysis (FPCA) approach to find the basis functions in Section \ref{ssec:fpca} and the estimation procedure for both factors and factor loadings in Section \ref{ssec:estimation}. Section \ref{sec:data} presents the dataset taken for the empirical study, followed by a description of the estimation results in Section \ref{sec:results}. Here, first, we interpret obtained factors and factor loadings and propose a time series model for low-dimensional factors in \ref{ssec:results_var}. Finally in Section \ref{ssec:dispbacktest} we propose and compare alternative dispersion strategy setups: a no hedge, a naïve approach and an advanced hedge. Section \ref{sec:conclusions} concludes.

\section{Correlation trading}\label{sec:corr}
\subsection{Average basket correlation}\label{ssec:corr}
In a basket of $N$ assets, correlation $\rho_{i,j}$ measures linear dependence between the $i$-th and the $j$-th asset return, $i,j \in \{1, \ldots, N\}$. Standard statistical analysis yields that the basket variance $\sigma^2_B$ can be decomposed as:

\begin{equation}\label{eq:baskvar}
\sigma^2_{B}=\sum_{i}w_i^2\sigma_i^2+\sum_{i}\sum_{j\neq i}w_iw_j\sigma_i\sigma_j\rho_{ij},
\end{equation}

where $\sigma^2_i$ denotes the variance of the $i$-th asset return and $w_i$ its weight in the basket.  Now, assuming that $\rho_{ij}$ is constant for every pair  $(i,j)$, one can imply the equicorrelation $\rho$  from (\ref{eq:baskvar}):

\begin{equation}\label{eq:eqcorr}
\rho=\frac{\sigma^2_{B}-\sum_{i}w_i^2\sigma_i^2}{\sum_{i}\sum_{j\neq i}w_iw_j\sigma_i\sigma_j}.
\end{equation}

Later we call $\rho$ a basket correlation or simply a correlation. The corresponding correlation matrix has all the off-diagonal elements equal to $\rho$ and thus offers several advantages. Firstly, plugging $\rho_{i,j}=\rho$ into  (\ref{eq:baskvar}) reproduces the basket variance $\sigma^2_{B}$. Secondly, if $-\frac{1}{N - 1}<\rho<1$ then the correlation matrix is positive semi-definite, \cite{Haerdle2015}. This property becomes particularly  important if $N$ is large. A closer look also reveals that (\ref{eq:eqcorr}) is in fact a nonlinear weighted average over all $\rho_{i,j}$ in the basket:

\begin{equation}\label{eq:eqcorr_waver}
\rho=\sum_{i}\sum_{j\neq i}c_{i,j}\rho_{i,j}
\end{equation}

with weights  $c_{i,j}$ defined by:

\begin{equation}\label{eq:eqcorr_waver_second}
c_{i,j}=\frac{w_iw_j\sigma_i\sigma_j}{\sum_{i}\sum_{j\neq i}w_iw_j\sigma_i\sigma_j}.
\end{equation}

\cite{Bourgoin2001} showed that if a correlation matrix is positive semi-definite, for sufficiently large baskets it holds that $0 \leq \rho \leq 1$. Using this property, maximum and minimum variances of a basket, $\sigma^2_{B, min}$ and $\sigma^2_{B,max}$ respectively, are defined as follows:

\begin{equation}\label{eq:baskvar_min}
\sigma^2_{B,min}=\sum_{i}w_i^2\sigma_i^2,
\end{equation}

\begin{equation}\label{eq:baskvar_max}
\sigma^2_{B,max}=\sum_{i}w_i^2\sigma_i^2+\sum_{i}\sum_{j\neq i}w_iw_j\sigma_i\sigma_j.
\end{equation}

$\sigma^2_{B,min}$ is achieved when $\rho=0$ that is when the assets in a basket are fully diversified. In the case of no diversification, one observes the maximal possible basket variance $\sigma^2_{B,max}$ corresponding to $\rho=1$.

Further we can rewrite $\rho$ by substituting (\ref{eq:baskvar_min}) and (\ref{eq:baskvar_max}) to (\ref{eq:eqcorr}):

\begin{equation}\label{eq:eqcorr_alternative}
\rho=\frac{\sigma^2_{B}-\sigma^2_{B,min}}{\sigma^2_{B,max}-\sigma^2_{B,min}}
\end{equation}

and obtain an additional interpretation as a measure for the degree of diversification, \cite{Skintzi2005}. In fact (\ref{eq:eqcorr_alternative}) shows how far $\sigma^2_{B}$ is from its minimal value $\sigma^2_{B,min}$ relative to the possible value range $\sigma^2_{B,max}-\sigma^2_{B,min}$, or in other words, how far the basket is  from the perfect diversification. High $\rho$ is the sign of a poorly diversified portfolio, which is typical for the market downturn, when asset prices simultaneously drop driving $\sigma^2_{B}$ up. It means diversification benefits disappear in times when they are needed most. To hedge against correlation risk investors look for derivative securities that offer higher payoffs (premia) when the correlation decreases. 

\begin{figure}[t!]
\begin{center}
\includegraphics[width=1\textwidth]{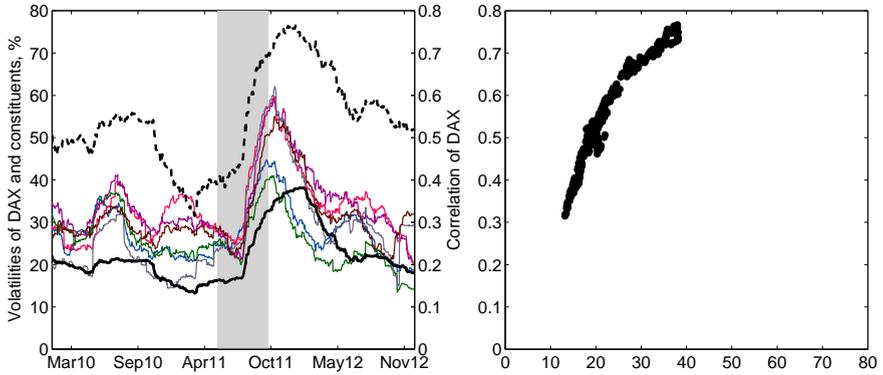}
\caption{Left panel: DAX correlation (\ref{eq:eqcorr}) - dashed, DAX volatility (\ref{eq:rv}) - solid black, volatility of DAX constituents Adidas, BMW, Siemens, Daimler, E.ON, Lufthansa volatilities (\ref{eq:rv}) - color, the stock market fall 2011 - shaded area. Right panel: scatter plot DAX volatility vs. correlation. Estimation period - from 20100104 to 20121228, estimation window - 3 months.\label{fig:DAXdrivers}}
\end{center}
\end{figure}

If a basket is constructed from the constituents of an equity index with weights equal to index weights, then the corresponding basket correlation would serve as a benchmark for a sector, an industry or a whole market average correlation. Figure \ref{fig:DAXdrivers} shows an example of the  DAX correlation together with the volatility of DAX and some of its components. Firstly, we see that the correlation and the volatility vary over time. Secondly, the volatility of the basket (DAX) is smaller than almost any individual volatility of its constituents, which illustrates the impact of the diversification effect on the portfolio risk. Finally, there is a clear linear dependence of the correlation of the basket and its volatility. However the strength of this dependence changes when the volatility exceeds a certain threshold. We investigate this phenomenon and propose a dataset correction scheme in Section \ref{sec:data}.

\subsection{Implied versus realized correlation}\label{ssec:mfic_vs_rc}

Based on (\ref{eq:eqcorr}) we conclude that the exposure to the basket correlation $\rho$ can be achieved by exposures to the variances of a basket $\sigma^2_B$ and its constituents, $\sigma^2_i$. Such trades can be realized via a combination of variance swaps. A variance swap is an over-the-counter contract opened at $t$, which at $t+\tau$ pays the difference between the variance cumulated over the life time of the swap $\sigma_{t+\tau}^2$ and the fixed pre-defined strike $\tilde{\sigma_{t}}^2(\tau)$:

\begin{equation}\label{eq:vswaps}
\left\{\sigma^2_{t+\tau}-\tilde{\sigma}^2_{t}(\tau)\right\}N_{var},
\end{equation}

where $N_{var}$ is the notional amount. Here and later $t$ and $\tau$ are given in fractions of a year.

The strike of a variance swap is the risk-neutral expectation at $t$ of the integrated variance from $t$ to $t+\tau$. It is also known as the model-free implied variance (MFIV), where ``model-free'' indicates that the expectation does not depend on the specification of the underlying price process, \cite{Britten-Jones2000}. MFIV can be approximated by a function of current option prices, \cite{Breeden1978}, \cite{Carr98towardsa}, \cite{Britten-Jones2000}, which has the following form

\begin{equation*}
\tilde{\sigma_{t}}^2(\tau)= {\mathop{\mbox{\sf E}}}_t^Q \left[\int^{t+\tau}_{t}\sigma^2(s)ds\right] = 
\end{equation*}
\begin{equation}\label{eq:mfiv}
\frac{2e^{r\tau}}{\tau}\left\{\int_0^{S_t}  \frac{P_t(K,\tau)dK}{K^{2}}+\int_{S_t}^\infty \frac{C_t(K,\tau)dK}{K^{2}}\right\},
\end{equation}

where $\mathop{\mbox{\sf E}}_t^Q$ expected value at $t$ under the risk-neutral measure $Q$, $P_t(K,\tau)$ $\left\{C_t(K,\tau)\right\}$ price at $t$ of put \{call\} with exercise price $K$ and time to maturity $\tau$, $S_t$ price of the asset in $t$, $r$ the annualized continuously compounded risk-free interest rate.

MFIV can be opposed to the implied variance $\widehat{\sigma}_t^2(\kappa,\tau)$, the square of the implied volatility (IV), which is obtained by solving
 
\begin{equation}\label{eq:iv}
V_t(\widehat{\sigma},\kappa,\tau)-\breve{V}_t(\kappa,\tau)=0,
\end{equation}

where $V_t$ is the theoretical (model) option price,  $\breve{V}_t$ option price taken from the market, $\displaystyle \kappa=\frac{K}{S_te^{r\tau}}$ moneyness of the option. IV, in comparison to MFIV, is a function of both $\kappa$ and $\tau$, meaning  that at every $t$ one recovers a cloud of points, which can be approximated by a surface, \cite{Cont2002}, \cite{Fengler2007}.

The floating leg of the variance swap, the realized variance (RV) of an asset from $t$ to $t+\tau$, can be computed from the time series of daily asset returns in different ways, depending on the contract specification. Here we use the most common following form

\begin{equation}\label{eq:rv}
\sigma_{t+\tau}^2={\tau}^{-1}\sum_{i=252t}^{252(t+\tau)}\left(\log \frac{S_i}{S_{i-1}}\right)^2.
\end{equation}

In \cite{Carr2009} $\sigma^2_{t+\tau}-\tilde{\sigma}^2_{t}(\tau)$ is referred to as the variance risk premium  (VRP), which is shown to be strongly negative for major US stock indexes over the sample period from January 1996 to December 2003. The negative sign indicates that investors are willing to pay extra to hedge themselves against possible future market turmoil. \cite{Bakshi2003}, who investigated the S\&P100 index and its largest constituents from 1991 to 1995, also found significant negative difference between realized and option implied volatilities for the average of 25 stocks and stressed that this difference is less pronounced than for the index. \cite{Vilkov2009} study each  S\&P100 constituent individually. Their $t$-test for $H_0$, that the sample means of RV and MFIV are equal, was not rejected for the majority of stocks in the sample from January 1996 to December 2003.

We check the same hypothesis on the German market for the sample period from 20100104 to 20121228 using the dataset described in Section \ref{sec:data}. Table \ref{table:t_test_DAX} in Appendix \ref{append:crp}  summarizes the results of a $t$-test for the null hypothesis that RV and MFIV are on average equal against the alternative RV$<$MFIV. $H_0$ is strongly rejected for the DAX index. For the DAX constituents the rejection rate decreases with the options' maturity $\tau$:  with $\tau=0.25$ (3 months) and $\tau=0.5$ (6 months) the $H_0$ cannot be rejected at a 5\% significance level for 8 out of 30 DAX constituents, with $\tau=1$ (1 year) for 13 constituents. Table \ref{table:t_test_DAX} in the  Appendix \ref{append:crp} reports the t-test results for these 13 stocks. In addition, Table \ref{table:mean_rv_mfiv_DAX} in the  Appendix \ref{append:vrp} report sample averages of RV and MFIV and their differences for all 30 DAX constituents. The latter are found to be negative for most of the stocks and for the DAX index.

\cite{Vilkov2009} interpret their $t$-test results as indirect evidence that a negative correlation risk premium (CRP) exists. To identify the existence of CRP in the DAX dataset we compute the model free implied correlation (MFIC) $\tilde{\rho}_{t}(\tau)$ from the MFIVs of DAX and its constituents and the realized correlation (RC) $\rho_{t+\tau}$  from the corresponding RV by applying (\ref{eq:eqcorr}):

\begin{equation}\label{eq:mfic}
\tilde{\rho}_{t}(\tau)=\frac{\tilde{\sigma}^2_{t,DAX}(\tau)-\sum_{i}w_i^2\tilde{\sigma}^2_{t,i}(\tau)}{\sum_{i}\sum_{j \neq i} w_i w_j \tilde{\sigma}_{t,i}(\tau)\tilde{\sigma}_{t,j}(\tau)},
\end{equation}

\begin{equation}\label{eq:rc}
\rho_{t+\tau}=\frac{\sigma^2_{t+\tau,DAX}-\sum_{i}w_i^2\sigma_{t+\tau,i}^2}{\sum_{i}\sum_{j\neq i}w_iw_j\sigma_{t+\tau,i}\sigma_{t+\tau,j}}.
\end{equation}

\begin{figure}[t!]
\begin{center}
\includegraphics[width=1\textwidth]{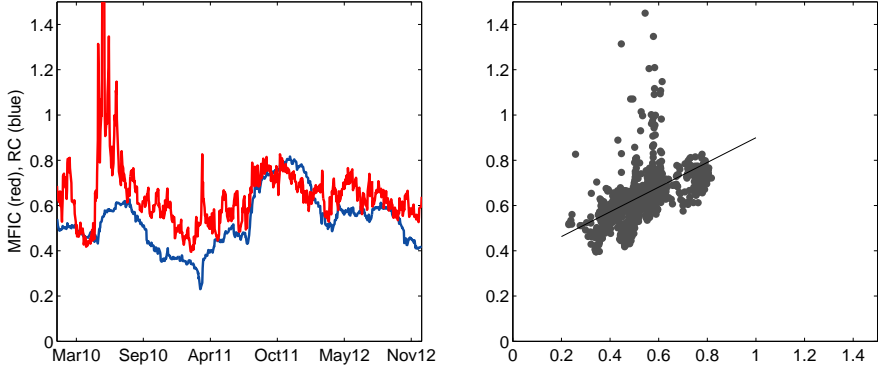}
\caption{Left panel: DAX $\rho_{t,\tau}$- blue, $\tilde{\rho}_{t}(\tau)$ - red, right panel: scatter plot of DAX $\rho_{t,\tau}$ (horizontal axis) vs $\tilde{\rho}_{t}(\tau)$ (vertical axis), for $t+0.25$ from 20100802 till 20120801.\label{fig:DAX_RC_MFIC_3m}}
\end{center}
\end{figure}

Figure \ref{fig:DAX_RC_MFIC_3m} plots the MFIC and the RC of DAX computed over the 3-month window and with 3 month maturity respectively ($\tau=0.25$). The $H_0:$ RC$=$MFIC of the $t$-test is strongly rejected. Using this finding and taking into account results in the literature we would expect the $\rho_{t+\tau}-\tilde{\rho}_{t}(\tau)$ (CRP) to be negative most of the time. One of the ways of exploiting this observation is to make a bet on the market correlation by entering a dispersion strategy.

\subsection{Dispersion strategy with variance swaps}\label{ssec:dispersion}

We study one of the variations of the dispersion strategy, which consists of selling the variance of the basket (DAX) and buying variances of basket constituents. 

The dispersion strategy can be implemented by taking a short position in the variance swap (\ref{eq:vswaps}) on an index and long positions in variance swaps on its constituents with notional amounts proportional to index weights. The payoff of a dispersion strategy at $t+\tau$ is then defined by

\begin{equation}\label{eq:disp_payoff1}
D_{t+\tau}=-\left\{\sigma^2_{t+\tau,B}-\tilde{\sigma}^2_{t,B}(\tau)\right\}+\sum_{i=1}^Nw^2_i\left\{\sigma^2_{t+\tau,i}-\tilde{\sigma}^2_{t,i}(\tau)\right\}.
\end{equation}

Then we apply (\ref{eq:eqcorr}) and rewrite (\ref{eq:disp_payoff1}) in the following form:

\begin{equation}\label{eq:disp_payoff2}
D_{t+\tau}=\tilde{\rho}_{t}(\tau)\sum_{i}\sum_{j \neq i} w_i w_j \tilde{\sigma}_{t,i}(\tau)\tilde{\sigma}_{t,j}(\tau)-\rho_{t+\tau}\sum_{i}\sum_{j \neq i} w_i w_j \sigma_{t+\tau,i}\sigma_{t+\tau,j}.
\end{equation}

Based on empirical findings described in Section \ref{ssec:mfic_vs_rc} we  assume $\tilde{\sigma}_{t,i}(\tau)\approx \sigma_{t+\tau,i}$ for each constituent stock and simplify the payoff (\ref{eq:disp_payoff2}), as follows

\begin{equation}\label{eq:disp_payoff_approx}
D_{t+\tau}\approx\sum_{i}\sum_{j \neq i} w_i w_j \tilde{\sigma}_{t,i}(\tau)\tilde{\sigma}_{t,j}(\tau)\left\{\tilde{\rho}_{t}(\tau)-\rho_{t+\tau}\right\},
\end{equation}

which illustrates that by entering the dispersion strategy one obtains exposure to $\rho_{t+\tau}-\tilde{\rho}_{t}(\tau)$, where the floating leg $\rho_{t+\tau}$ is computed with (\ref{eq:rv}) and (\ref{eq:eqcorr}) at expiry, and the fixed leg $\tilde{\rho}_{t}(\tau)$ is a function of variance swap strikes (\ref{eq:mfiv}). Test results described in Section \ref{ssec:mfic_vs_rc} suggest that we should, on average, expect  $\rho_{t+\tau}-\tilde{\rho}_{t}(\tau)<0$. It also means  the dispersion strategy with payoff $D_{t+\tau}$ on average would have a profit. However, as one can see in Figure \ref{fig:DAX_RC_MFIC_3m}, there might be days when $\rho_{t+\tau}-\tilde{\rho}_{t}(\tau) \geq 0$. In order to hedge against these potential losses one needs a forecast of the floating leg of the dispersion strategy.

Another possible modification of the dispersion trading strategy does not involve trading on the OTC market and can be implemented with standardized market instruments, puts and calls. The strategy consists in selling index option straddles and purchasing straddles in options on index components. The forecast of the implied correlation surface can provide the insight into the relative cost of index options compared to the price of options on individual stocks that comprise the index. In comparison to the single historical or implied volatility forecast, usually used for this purpose, the correlation surface can provide information for trading options on the whole maturity spectrum. Which means one can buy straddles with different strikes, depending on the implied correlation forecast.

\section{Modeling and forecasting correlation dynamics}\label{sec:cormod}

To determine the amount of hedge for $D_{t+\tau}$ we model the implied correlation (IC) and use the forecast to approximate the floating leg of the dispersion strategy $\rho_{t+\tau}$. By applying (\ref{eq:eqcorr})  to IV of a basket $\widehat{\sigma}_{t,B}(\kappa, \tau)$ and its $N$ constituents $\widehat{\sigma}_{t,i}(\kappa, \tau)$, $i \in \{1, \ldots,N\}$, every $t$ we obtain the IC surface (ICS):

\begin{equation} \label{eq:ics}
\widehat{\rho}_t(\kappa,\tau)=\frac{\widehat{\sigma}^2_{t,B}(\kappa,\tau)-\sum_{i}w_i^2\widehat{\sigma}_{t,i}^2(\kappa,\tau)}{\sum_{i}\sum_{j\neq i}w_iw_j\widehat{\sigma}_{t,i}(\kappa,\tau)\widehat{\sigma}_{t,j}(\kappa, \tau)}.
\end{equation}

\begin{figure}[t!]
\begin{center}
\includegraphics[width=1\textwidth]{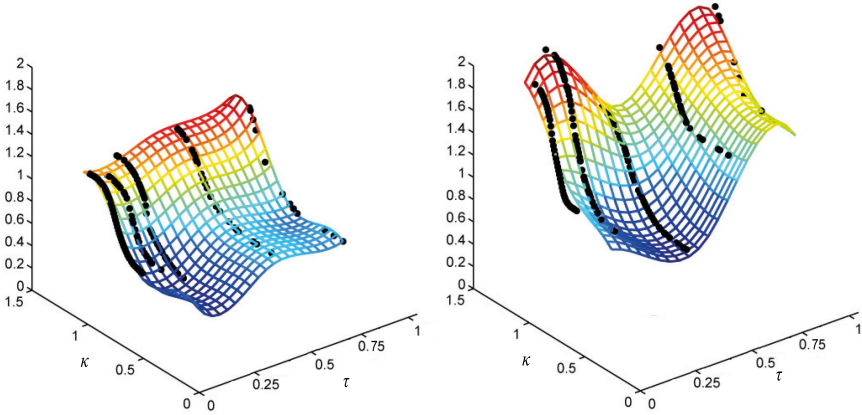}
\caption{ICS implied by prices of DAX options traded on the 20111209, 20120710, surfaces recovered by the Nadaraya-Watson smoothing}
\label{fig:sp500_ICS}
\end{center}
\end{figure}

Figure \ref{fig:sp500_ICS} displays $\widehat{\rho}_t(\kappa,\tau)$ in different trading days:  20111209, 20120710. Due to the specific option data structure, every day one observes a ``cloud of strings'' that visually resembles a surface and can be recovered by applying nonparametric smoothing. One can clearly see that surfaces have shape similarities, but vary in levels, slopes and curvatures. Thus they may be treated as daily realizations of a random function. In addition one can observe that the strings do not have fixed spacial locations. In order to model the dynamics of such  a complicated multi-dimensional object we apply the DSFM that reduces the dimensionality of the problem and allows the ICS to be studied in a conventional time-series context.

\subsection{Model Characterization}\label{ssec:model}
At every day $t$ one observes ICs $\widehat{\rho}(\kappa_{t,j},\tau_{t,j})$, $t=1,\ldots,T$, $j=1,\ldots,J_t$, where $j$ is the index of observations and $J_t$ the total number of observations at day $t$.  Prior to introducing the model we exclude the case of a fully undiversified basket, with $\widehat{\rho}=1$, from the analysis and apply a variance stabilizing transformation. Fisher's Z-transformation (\cite{Haerdle2015}) gives:

\begin{equation}\label{eq:fisher_transf}
T(u) \stackrel{\operatorname{def}}{=}\frac{1}{2}\log\frac{1+u}{1-u}
\end{equation}

with $Y_{t,j}\stackrel{\operatorname{def}}{=} T\left\{\widehat{\rho}(\kappa_{t,j},\tau_{t,j})\right\}$.

Our aim is to model the dynamics of  $\left\{ (Y_{t,j},X_{t,j}), 1 \leq t \leq T, 1 \leq j \leq J_t\right\}$, where $X_{t,j}=(\kappa_{t,j},\tau_{t,j})$. The technique we employ allows us to reduce the dimensionality and to simultaneously study the dynamics of  $Y_t$ by approximation through an $L$-dimensional object with $L<<J$. The DSFM, first introduced by \cite{Fengler2007} in an application to IV surface dynamics, and then extended by \cite{Park2009} and \cite{Song2014} has these desired properties.

The basic idea is to approximate  $\mathop{\mbox{\sf E}}(Y_{t}|X_{t})$ by the sum of $L+1$ smooth basis functions $m\stackrel{\operatorname{def}}{=}\left\{m_{0},\ldots,m_{L}\right\}^{\top}$ (factor loadings) weighted by time dependent coefficients  $Z_t\stackrel{\operatorname{def}}{=}(1,Z_{t,1},\ldots,Z_{t,L})^{\top}$ (factors):

\begin{equation}\label{eq:model_spec}
Y_{t,j}=m_0(X_{t,j})+\sum_{l=1}^LZ_{t,l}m_l(X_{t,j})+\varepsilon_{t,j}.
\end{equation}

In representation (\ref{eq:model_spec}) $m$ are chosen data driven and do not have a particular (parametric) form. 

Here two important remarks are appropriate. Firstly, the unknown basis functions $m$ must be estimated. \cite{Fengler2007} estimate both $m$ and $Z_t$ iteratively using kernel smoothing techniques, \cite{Park2009} approximates $m$ by tensor B-splines basis functions weighted by a coefficients matrix.  Here we employ a functional principal component analysis (FPCA) approach that will be described in Section \ref{ssec:fpca}. The nonparametric estimation procedure that we use is introduced in Section \ref{ssec:estimation}; the basics of this technique can be found in \cite{Song2014}.

The second issue is the estimation of the latent factors $Z_t$. Having the data-driven basis $\widehat m_l$ in hand, we can estimate daily factors by the ordinary least squares (OLS) method. Afterwards one fits the econometric model to $\widehat Z_t$, as it was done by \cite{Cont2002} and \cite{Hafner2004}, who fitted AR(1) to every $Z_{t,l}$, $l \in \{1, \ldots, L\}$, or in \cite{Fengler2007} who considered a multivariate VAR(2) process.

\subsection{Correlation surface with FPCA}\label{ssec:fpca}

We approximate the ICS by the sum of orthogonal functions. By doing so we involve the FPCA theory by looking at the ICS as a stationary random function $f:\mathbb{R}^2\rightarrow\mathbb{R}$. 

Let  $\mathscr{J}=[\kappa_{min},\kappa_{max}]\times[\tau_{min},\tau_{max}]$  the range of possible values of  $\kappa_{t,j}$ and $\tau_{t,j}$. We introduce $(\rho_t)$, $t \in \{1,\ldots ,T\}$, the sample of i.i.d. smooth random functions (surfaces). Every $\rho_t$  is a smooth map $\rho_t: \mathscr{J} \rightarrow\mathbb{R}$ and satisfies $\int_{\mathscr{J}}\mathop{\mbox{\sf E}}(\rho_t^2)<\infty$. Also for every $\rho_t$ we assume a well-defined mean function $\mu(u)=\mathop{\mbox{\sf E}}\left\{\rho_t(u)\right\}$ and an existence of a covariance function $\psi(u,v)=\mathop{\mbox{\sf E}}\left[\left\{\rho_t(u)-\mu(u)\right\}\left\{\rho_t(v)-\mu(v)\right\}\right] $. With $\phi(u,v)=\mathop{\mbox{\sf E}}\left\{\rho_t(u)\rho_t(v)\right\} $ the covariance function can be expressed as

\begin{equation}\label{eq:cov_coef}
\psi(u,v)=\phi(u,v)-\mu(u)\mu(v),
\end{equation}

which can be also interpreted as a covariance coefficient of two points on the surface with coordinates $u$ and $v$ $\in \mathscr{J}$. 
Since (\ref{eq:cov_coef}) is a symmetric positive definite function we can use it as a nucleus of the integral transform, performed by the linear operator. Define the covariance operator $\Gamma$:

\begin{equation}\label{eq:cov_op}
(\Gamma f)(u)=\int_\mathscr{J} \psi(u,v)  f(v) dv
\end{equation}

that transforms $f$ into $(\Gamma f)$. $\Gamma$ is a symmetric positive operator with orthonormal eigenfunctions  $ \left\{\gamma_j\right\}_{j=1}^{\infty}$,  $\gamma_j: \mathscr{J} \rightarrow\mathbb{R}$,  and associated eigenvalues $ \left\{\lambda_j\right\}_{j=1}^{\infty}$ with $\lambda_1\geq \lambda_2\geq  \ldots \geq  0$.  Now we can express (\ref{eq:cov_coef}) in terms of eigenfunctions and eigenvalues of the covariance operator $\Gamma$ by applying Mercer's theorem, e.g. \cite{Indritz1963}:

\begin{equation}\label{eq:mercer}
\psi(u,v)=\sum_{j=1}^{\infty}\lambda_j\gamma_j(u)\gamma_j(v).
\end{equation}

Taking eigenfunctions $ \left\{\gamma_j\right\}_{j=1}^{\infty}$ as a basis, we represent $\rho_t(u)-\mu(u)$ as a generalized Fourier series with coefficients given by $\zeta_{tj}=\int_\mathscr{J} \left\{\rho_t(u)-\mu(u)\right \}\gamma_j(u) du$ called the $j$-th principal component score with $\mathop{\mbox{\sf E}}(\zeta_{tj})=0$, $\mathop{\mbox{\sf E}}(\zeta_{tj}^2)=\lambda_j$ and $\mathop{\mbox{\sf E}}(\zeta_{tj}\zeta_{ik})=0$ for $j\neq k$, \cite{Silverman2010}. Thus one may rewrite $\rho_t(u)-\mu(u)$ in the Karhunen-Lo\`{e}ve form:

\begin{equation}\label{eq:karhunen-loeve}
\rho_t(u)-\mu(u)=\sum_{j=1}^\infty\zeta_{tj}\gamma_j(u).
\end{equation}

Here $\zeta_{tj}$ indicates how strong the influence of the $j$-th basis function on the shape of the $t$-th surface is. The higher the score, the closer the shape of $\rho_t$ resembles the shape of the $j$-th eigenfunction. 

In practice one needs to take $L$ eigenfunctions to replace the infinite sum in (\ref{eq:karhunen-loeve})  by the finite sum of $L$ basis functions, corresponding to the highest eigenvalues. One calls  $ \left\{\gamma_j\right\}_{j=1}^{L}$ the empirical orthonormal basis, \cite {Silverman2010}. In the next Section we discuss the estimation procedure for  $ \left\{\gamma_j\right\}_{j=1}^{L}$ as well as criteria for the $L$ selection.

\subsection{Estimation Algorithm}\label{ssec:estimation}

In model (\ref{eq:model_spec}) both $Z_t$ and $m$ must be estimated. We do that in two steps.

At the {\bf first step} we estimate the covariance operator introduced in Section (\ref{ssec:fpca}) and take $\widehat  \mu$ as $\widehat m_0$ and $\widehat  \gamma_l$ as  $\widehat  m_l$, $l \in \{1,\ldots, L\}$ .

The covariance function (\ref{eq:cov_coef}) is estimated as described in \cite{Yao2005} and \cite{Hall2006}. The procedure consists in least-squares fitting of two local linear models, for $\widehat \mu$ and $\widehat \psi$.

Given $u\in \mathscr{J}$ we choose $(\widehat a_{\mu},\widehat b_{\mu})=(a_{\mu}, b_{\mu})$ to minimize

\begin{equation}\label{eq:1est_step}
\sum^T_{t=1}\sum^{J_t}_{j=1} \{Y_{t,j}-a_{\mu}-b_{\mu}(u-X_{t,j}) \}^2\mathcal{K}_{h_\mu}\left(X_{t,j}-u\right) ,
\end{equation}

and take $\widehat\mu(u)=\widehat a_{\mu}$. Then, given $u,v\in \mathscr{J}$ we choose $(\widehat a_{\phi},\widehat b_{\phi,1},\widehat b_{\phi,2})=(a_{\phi},b_{\phi,1},b_{\phi,2})$ to minimize

\begin{align}\label{eq:2est_step}
\sum^T_{t=1}\sum_{j,k:1\leq j \neq k\leq J_t} \{Y_{t,j}Y_{t,k}- a_{\phi}-b_{\phi,1}(u-X_{t,j})-b_{\phi,2}(v-X_{t,k}) \}^2\\
\nonumber
\times\mathcal{K}_{h_\phi}\left(X_{t,j}-u\right)\mathcal{K}_{h_\phi}\left(X_{t,k}-v\right),
\end{align}

and take $\widehat\phi(u,v)=\widehat a_{\phi}$.

Here $\mathcal{K}_{h}$ denotes the two-dimensional product kernel, $\mathcal{K}_{h}(\bar{q})=k_{h_1}(\bar{q}_1)\times k_{h_2}(\bar{q}_2)$, $h=(h_1,h_2)^{\top}$, based on one-dimensional $k_{h}(\bar{q})=h^{-1}k(h^{-1}\bar{q})$. For our application we selected the quartic kernel, where $k(\bar{q})=15/16(1-\bar{q}^2)^2$ for $|\bar{q}|<1$ and $0$ otherwise. For both (\ref{eq:1est_step}) and (\ref{eq:2est_step}) kernel bandwidths ${h_\mu=(h_{\mu,1},h_{\mu,2})^{\top}}$ and ${h_\phi=(h_{\phi,1},h_{\phi,2})^{\top}}$ are to be selected. The procedure is described in Appendix \ref{apsec:badnwidth}. Figure \ref{fig:mu_u_and_data_DAX2010} shows an example of $\widehat\mu(u)$ estimated using the dataset described in Section \ref{sec:data} for a sub-sample from 20100802 to 20110801.

Finally, having estimates $\widehat \mu(u)$ and $\widehat \phi(u,v)$, we compute $\widehat \psi(u,v)$ using (\ref{eq:cov_coef}) and take its $L$ eigenfunctions corresponding to the largest eigenvalues as $\widehat m_{l}$, $l \in \{1,\ldots,L\}$.
Parameter $L$ is chosen in such a way that the selected eigenfunctions explain the large share of variability in the original data. It is also necessary to mention that $\widehat \psi(u,v)$ is a matrix of a very large dimensionality. To obtain its consistent estimator, suitable for further spectral decomposition, various matrix regularization techniques can be used., e.g. banding as in \cite{Bickel2008}, thresholding in \cite{Bickel2008a}, eigenvalues cleaning as in \cite{LaurentLaloux1999} and factor models described in  \cite{Fan2008}. We use the latter in this step.

\begin{figure}[t!]
\begin{center}
\includegraphics[width=1\textwidth]{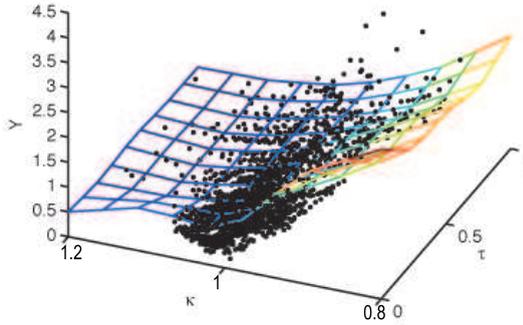}
\caption{Mean function $\widehat\mu(u)$ of the DAX ICS with corresponding data points, estimated from 20100802 to 20110801 with $h_\mu=(h_{\mu,1},h_{\mu,2})^{\top}=(0.12,0.17)^{\top}$}
\label{fig:mu_u_and_data_DAX2010}
\end{center}
\end{figure}

In the {\bf second step} using $\widehat m$ we obtain the estimates $\widehat Z_t=(1,\widehat Z_{t,1},\ldots,\widehat Z_{t,L})^{\top}$ as minimizers of the following least squares criterion:

\begin{equation}\label{eq:ols}
\widehat Z_{t}=\operatorname{arg}\,\underset{Z_{t}} {\operatorname{min}} \sum_{t=1}^T \sum_{j=1}^{J_t} \left \{Y_{t,j} - 
Z_{t}^{\top}
\widehat m(X_{t,j}) \right \}^2.
\end{equation}

\section{Data}\label{sec:data}

\begin{table}[t!]
\begin{center}
\begin{tabular}{*{2}{c}*{7}{c}}\hline \hline
			&																	&	Min.		&	Max.		&	Mean		&	Median	&	Stdd.		&	Skewn.	&	Kurt\\
\hline
IC		&	$\kappa$												&	0.8000	&	1.2000	&	 0.9825	&	0.9825	&	0.0986	&	~0.0690	&	~2.0661\\
			&	$\tau$													&	0.0274	&	0.9671	&	0.2442	&	0.1753	&	0.1979	&	~1.3717	&	~4.3941\\
			&	$\widehat \rho_t(\kappa,\tau)$	&	0.0587	&	0.9998	&	0.6150	&	0.6290	&	0.1566	&	-0.2739	&	~2.6115\\
\hline
MFIC	&	$\tilde{\rho}_{t}(0.083)$				&	0.3895	&	0.4860	&	0.6061	&	0.6193	&	0.0834	&	0.0696	&	~0.1957\\
			&	$\tilde{\rho}_{t}(0.25)$				&	0.4446	&	0.9795	&	0.6549	&	0.6573	&	0.0850	&	0.0613	&	~0.1631\\
			&	$\tilde{\rho}_{t}(0.5)$					&	0.4997	&	1.4730	&	0.7037	&	0.6953	&	0.0866	&	1.8188	&	~0.1305\\
			&	$\tilde{\rho}_{t}(1)$						&	0.5611	&	1.0851	&	0.7496	&	0.7422	&	0.0905	&	0.7764	&	~0.6788\\
\hline
RC		&	$\rho_{t+0.083}$								&	0.1754	&	0.8955	&	0.5373	&	0.5013	&	0.1331	&	0.5221	&	-0.2154\\
			&	$\rho_{t+0.25}$									&	0.2774	&	0.8149	&	0.5566	&	0.5363	&	0.1192	&	~0.2489	&	-0.8083\\
			&	$\rho_{t+0.5}$									&	0.3794	&	0.7343	&	0.5759	&	0.5713	&	0.1053	&	-0.0243	&	-1.4012\\
			&	$\rho_{t+1}$										&	0.4312	&	0.6581	&	0.5924	&	0.6050	&	0.0522	&	-1.2443	&	~0.9875\\
\hline\hline
\end{tabular}
\caption{Summary statistics: IC data computed from the DAX index and constituents options over the period from 20090803 to 20120801 including the 1 year estimation period (3 years, 770 trading days, 135 obs./day). MFIC computed from daily variance swaps rates. RC computed from daily stock returns from 20100802 to 20120801 (2 years, 515 trading days). The figures are given after filtering and data preparation.}
\label{table:summary}
\end{center}
\end{table}

We study the dispersion strategy over the two year sample period from 20100802 to 20120801 on the German market represented by the DAX basket. The basket is composed of 23 stocks, constituents of DAX, with the most liquidly traded options and weights proportional to the current market capitalization. To model the dynamics of the IC and to construct the dispersion trade we operate with three main variables representing different correlation estimates. MFIC, RC, and IC. The datasets are described in Table \ref{table:summary}. 

The \textit{MFIC dataset} contains daily series of MFICs with maturities 0.083, 0.25, 0.5 and 1 years computed  via (\ref{eq:mfic}) from variance swap rates given by Bloomberg as a discrete approximation of (\ref{eq:mfiv}).

The \textit{RC dataset} contains daily series of RCs computed with (\ref{eq:rv}) and (\ref{eq:rc}) from the Bloomberg end-of-day stock prices over estimation windows 0.083, 0.25, 0.5 and 1 years. 

The \textit{IC dataset} is constructed using out-of-the-money (OTM) DAX and single stock options from the EUREX database. To estimate the DSFM model and produce forecasts for the sample period the dataset covers one additional year from 20090803 to 20100730. The dataset is transaction-based, meaning every trade is registered with the date it occurred, expiry date, underlying ticker,  exercise price (strike) and settlement price. To obtain IV from option prices via (\ref{eq:iv}) we distinguish between index and single stock options. For index options, which have the European type of option payoff, the Black-Sholes (BS) model is used. To account for dividends and early execution in options on single stocks (American payoff) we use binomial trees, \cite{Cox1979}, and bisection algorithm. Other necessary model parameters, such as stock prices, index levels, dividend amounts for constituent stocks, interest rates and stock market capitalization are taken from the Bloomberg database. As a risk free rate proxy we take daily values of EURIBOR (Euro Interbank Offered Rate) with 1 week up to 1 year maturities and use linear interpolation to obtain values for required option $\tau$. We use the most liquid segment of data with $\kappa$ ranging from $0.8$ to $1.2$ and $\tau$  from 10 days to 1 year. Options outside of this range are excluded from the data set due to the poor data quality, which does not allow to recover implied volatility surfaces for the DAX and all constituents and to compute implied correlation on a daily basis. Figure \ref{fig:sp500_ICS} in Section \ref{sec:cormod} shows an example of the ICS plotted using the entire available option data, including options outside of the $\tau$-range from $0.8$ to $1.2$, for two selected ``rich with data'' days (20111209 and 2012071).  As one can see, some correlations observed in Figure \ref{fig:sp500_ICS} are more extreme in comparison to the values in Table \ref{table:summary}.  The plots show the nature of the implied correlation estimate, which is not necessarily observed in a range from 0 to 1. Those days reveal the possibility of a so-called ``volatility arbitrage''. Having in mind the empirical findings described in Section \ref{ssec:mfic_vs_rc}, stating that the VRP of an index is much more pronounced then of constituents, one might take a short position in a too-expensive delta-hedged index option, when the implied correlation is considerably higher than 1. 

Options from original EUREX dataset are not given on a regular $(\kappa, \tau)$-grid, required in (\ref{eq:ics}).  In the $\tau$-dimension, maturities are standardized by market regulation, so every $t$ one can find several $\tau_t$, similarly for the index and for all constituents. However, in $\kappa$-dimension one needs to interpolate. At every $t$ we use the original $(\kappa_t, \tau_t)$ grid of the index and linearly interpolate IVs of all constituents to obtain values corresponding to this grid. To avoid  computational problems with a highly skewed empirical distribution of $(\kappa_t, \tau_t)$, we transform the initial space $[0.8,1.2]\times [0.03,1]$  to $[0,1]^2$ using an empirical distribution function. Also, we remove options with extremely high IVs (larger than 50\%) considering them the misprints in trade registration and finally use (\ref{eq:ics}) to obtain IC, which produces, on average, 135 observations per day.

Figure \ref{fig:DAXdrivers} shows there is a linear dependence between basket correlation and volatility. We check this finding in the RC dataset for different estimation windows and in IC dataset for different maturities. The RC data allows for the identification of a breakpoint, a threshold, after which the strength of the dependence changes, Appendix \ref{apsec:breakpoints}. This phenomenon is persistent over different estimation windows. The IC dataset does not show any clear change in correlation/volatility dependence. Since the IC is used to obtain a forecast of a floating leg of the dispersion strategy, which is RC, we propose making a regime dependent correction of the IC dataset as described in Appendix \ref{append:switchpoint}.

\section{Empirical results}\label{sec:results}
\subsection{Estimation Results and Factor Modeling}\label{ssec:results_var}
\begin{figure}[t!]
\begin{center}
\includegraphics[width=1\textwidth]{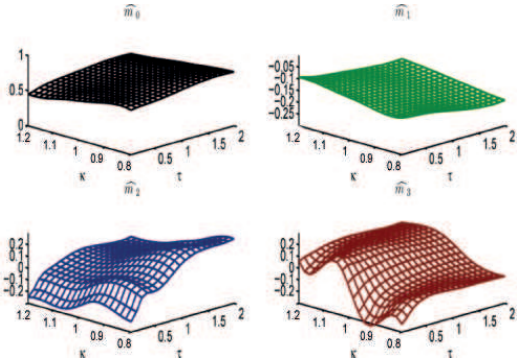}
\caption{Factor loadings $\widehat m_{0},\widehat m_{1},\widehat m_{2},\widehat m_{3}$ estimated from 20090803 to 20100730}
\label{fig:BASIS_FUNCTIONS_DAX_20120214-20120801}
\end{center}
\end{figure}

Using the IC dataset described in Section \ref{sec:data} we estimate the DSFM model for three non-overlapping sub-samples 20090803 - 20100730 (the 1st year), 20100802 - 20110729 (the 2nd year), 20110802 - 20120801 (the 3rd year), and for the entire sample 20090803 - 20120801. All sub-samples include particularly volatile periods caused by the stock market falls in May 2010, ``Flash Crash 2010'', and a more pronounced drop in August 2011.

\begin{figure}[t!]
\begin{center}
\includegraphics[width=1\textwidth]{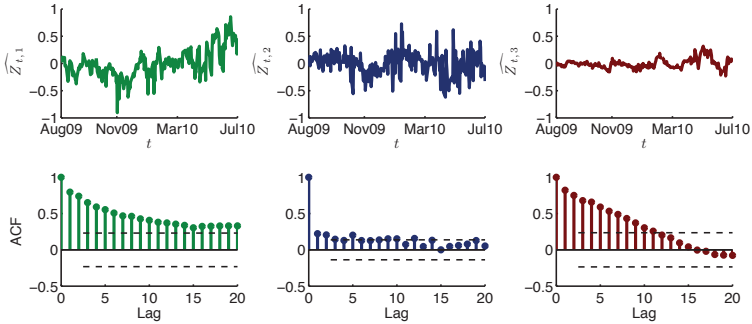}
\caption{Driving factors of the DAX ICS $\widehat Z_{t,1},\widehat Z_{t,2},\widehat Z_{t,3}$ and ACF up to the 20th lag from 20090803 to 20100730}
\label{fig:Zs_DAX_20120214-20120801}
\end{center}
\end{figure}

An example of an estimation over the 1st sample year common factor loadings $\widehat m_{0},\widehat m_{1},\widehat m_{2},\widehat m_{3}$ and the daily time series of factors $\widehat Z_{t,1},\widehat Z_{t,2},\widehat Z_{t,3}$ is given in Figure \ref{fig:BASIS_FUNCTIONS_DAX_20120214-20120801} and Figure \ref{fig:Zs_DAX_20120214-20120801}. Now the modeling task is simplified to the low-dimensional analysis of factor series. We fit the VAR model of order $p$ for $\widehat Z_{t,1},\widehat Z_{t,2},\widehat Z_{t,3}$. Before proposing a proper VAR specification, we check if $\widehat Z_{t}$ has characteristics that violate assumptions for linear multiple time series models. We perform the augmented Dickey-Fuller (ADF) test to check each $\widehat Z_{t,1},\widehat Z_{t,2},\widehat Z_{t,3}$ for stationarity, Appendix \ref{apsec:nlags}. For $\widehat Z_{t,2}$ in sub-sample 20100802 - 20110729 we cannot reject the hypothesis of a unit root, so we use its first differences instead. Then we define the appropriate number of lags, or order $p$, by computing  Akaike's information criterion (AIC), Schwarz's Bayesian information criterion (SBIC), and the Hannan and Quinn information criterion (HQIC) values, Appendix \ref{apsec:nags}.  An * appearing next to the test statistics indicates the optimal lag. Except for the sub-sample 20110802 - 20120801, the test statistics suggest $p=2$, so we make a choice in favor of this specification. The estimation results are summarized in Appendix \ref{apsec:DAX_VAREST}.  We also conducted a portmanteau (Q) test for the null hypothesis that a series of residuals exhibits no autocorrelation. The test does not indicate the presence of a serial correlation. 

Based on the results, we can distinguish the influence of each factor on the time evolution of the ICS. The first factor can be interpreted as level, the second as maturity and  the third as a moneyness effect. The relative size of the largest eigenvalues of (\ref{eq:cov_coef}) suggest that $\widehat m_{1}$ is capable of capturing the biggest share of the surface variability. The variation captured by the second $\widehat m_{2}$ has a smaller influence, since it is only responsible for the surface shape transformation in the $\tau$ dimension. Finally, since the variation of the ICS in the $\kappa$ dimension is relatively small, the $\widehat m_{3}$ has a smaller impact, which is also reflected in the $\widehat Z_{t,3}$ series.

The forecast of $\widehat Z_{t,1},\widehat Z_{t,2},\widehat Z_{t,3}$ modeled with VAR(2) together with estimated fixed $\widehat m_{0}$, $\widehat m_{1}$, $\widehat m_{2}$, $\widehat m_{3}$ give a forecast of the ICS.
\subsection{Backtesting the dispersion strategy}\label{ssec:dispbacktest}

Here we show that using the correlation forecast one can improve the original dispersion strategy (\ref{eq:disp_payoff1}) and test it empirically over the 2-years sample period  20100801 - 20120802. We compare the payoff of the strategy without hedging with the  \textit{naïve  hedging strategy} and  propose its improvement, the \textit{advanced strategy}.

To obtain the value of the naïve hedge position to be held over $\Delta t$ days from $t+\tau-\Delta t$ till $t+\tau$ 
we make a $\Delta t$-days ahead DSFM forecast $\widehat{\rho}_{t+\tau}(1,t+\tau)$ and use it as $\rho_{t+\tau}$ in (\ref{eq:disp_payoff1}). Thus the size of the position is defined by

\begin{equation}\label{hedge}
D^h_{t+\tau}=\sum_{i}\sum_{j \neq i} w_i w_j \tilde{\sigma}_{t,i}(\tau)\tilde{\sigma}_{t,j}(\tau)\left\{\tilde{\rho}_{t}(\tau)-\widehat{\rho}_{t+\tau}(1,t+\tau)\right\}.
\end{equation}

The corresponding relative hedging error is given by

\begin{equation}\label{herror}
\varepsilon^h_{t+\tau}=\frac{D^h_{t+\tau}-D_{t+\tau}}{D_{t+\tau}}=-\frac{\widehat{\rho}_{t+\tau}(1,t+\tau)-\rho_{t+\tau}}{\tilde{\rho}_{t}(\tau)-\rho_{t+\tau}},
\end{equation}

where $\varepsilon^h_{t+\tau}<0(>0)$ means that the hedge (\ref{hedge}) under-(over-)estimates the actual position (\ref{eq:disp_payoff1}). Table \ref{table:naive} gives summary statistics for the (\ref{herror}) over the studied sample period for 3 trades with four different maturities: 0.083, 0.25, 0.5 and 1 years. The statistic includes 515 trades originated every day and expired over the given 2 year sample period, $\Delta t$ is one day.

\begin{table}[t!]
\begin{center}
\begin{tabular}{c|*{7}{c}}\hline \hline
$\tau$		&	Min.			&	Max.	&	Mean.		&	Median	&	Stdd. 	&	Skew. 		&	Kurt.\\
\hline
$0.083$		&-108.04	&	72.30	&	-1.14	&	-0.71	&	~8.00	&	~-6.61	&	100.49\\
$0.25~$		&-255.48	&	49.53	&	-1.20	&	-0.41	&	11.49	&	-17.58	&	372.33\\
$0.5~~$		&-216.04	&	32.78	&	-0.74	&	-0.30	&	~9.37	&	-18.66	&	425.86\\
$1~~~~$		&~-64.84	&	76.59	&	-0.01	&	-0.38	&	~7.47	&	~~2.74	&	~46.85\\
\hline\hline
\end{tabular}
\end{center}
\caption{Performance of naïve hedge, summary statistics for $\varepsilon^h_{t+\tau}$ from 20100101 to 20120801}
\label{table:naive}
\end{table}

The improved version of the strategy uses the DSFM forecast $\widehat{\rho}_{t+\tau}(1,t+\tau)$ as a trigger which defines whether one should hedge or not. If $\widehat{\rho}_{t+\tau}(1,t+\tau)\geq \tilde{\rho}_{t}(\tau)$ (DSFM predicts loss in dispersion strategy), takes an offsetting (with negative sign) position  in (\ref{hedge}); if $\widehat{\rho}_{t+\tau}(1,t+\tau) < \tilde{\rho}_{t}(\tau)$   (DSFM predicts gain in dispersion  strategy), do not hedge. Thus we can write the payoff of the advanced strategy at $t+\tau$ as follows:

\begin{equation}\label{disp_adv}
D^{adv}_{t+\tau} = \left\{ \begin{array}{l l l}
D_{t+\tau}-D^{h}_{t+\tau}	&\mathrm{, if }& \widehat{\rho}_{t+\tau}(1,t+\tau)\geq \tilde{\rho}_{t}(\tau)\\
D_{t+\tau} 								&\mathrm{, if }& \widehat{\rho}_{t+\tau}(1,t+\tau)<\tilde{\rho}_{t}(\tau).\\
\end{array}\right.
\end{equation}

Since a variance swap contract costs nothing to initiate (we ignore transactions costs), the presented series of daily payoffs correspond to daily P\&L of the hypothetical trade where swaps expire daily over the whole period from 20100801 till 20120802. We compare the cash flows from three strategies. As one can see in Table \ref{table:cashflows}, the advanced strategy outperforms the other two by having the smallest maximal losses, highest maximal gains ($\tau=0.25,0.5$) and the highest (second highest for $\tau=1$) average payoff over the studied sample period.

\begin{table}[t!]
\begin{center}
\begin{tabular}{c|c|c*{4}{c}}\hline \hline
Strategy													 & $\tau	$		&	Min.		&	Max.	&	Mean.	&	Stdd.\\
\hline
$D_{t+\tau}$ 												& 0.083			&-1502.58	&	\textit{1080.23}	&	~87.09	&	356.94\\
(no hedge)												 & 0.25~			& -1531.94	&	\textit{1282.31}	&	101.92	&	440.54\\
																	 & 0.5~~			& -1270.90	&	\textit{1301.28}	&	136.91	&	456.75\\
																	 & 1~~~~			& ~-872.76	&	\textit{~760.92}	&	\textit{134.26}	&	299.01\\
\hline 
$D_{t+\tau}-D^h_{t+\tau}$ 					& 0.083			& -3237.72	&	617.40	&	15.35	&	\textit{203.09}\\
(naïve hedge)										& 0.25~			& -1726.53	&	413.28	&	35.90	&	\textit{110.14}\\
																	 & 0.5~~			& -1301.47	&	344.91	&	41.13	&	\textit{~91.91}\\
																	 & 1~~~~			& ~-914.27	&	327.03	&	79.62	&	\textit{~93.14}\\
\hline
$D^{adv}_{t+\tau}$									& 0.083			& \textit{-1375.99}	&	1011.38	&	\textit{100.93}	&	256.50\\
(advanced hedge)								 & 0.25~			& \textit{-1137.79}	&	\textit{1282.31}	&	\textit{195.09}	&	248.41\\
																	 & 0.5~~			& \textit{~-760.85}	&	\textit{1301.28}	&	\textit{231.35}	&	281.66\\
																	 & 1					& \textit{~-367.89}	&	~623.38	&	123.04	&	190.80\\
\hline\hline
\end{tabular}
\end{center}
\caption{Summary statistics for payoffs $D_{t+\tau}$ (no hedge), $D_{t+\tau}-D^h_{t+\tau}$ (naïve hedge), $D^{adv}_{t+\tau}$ (advanced hedge) from 20100101 to 20120801, $\Delta t$ is one day, best results (highest min, max, mean and smallest stdd.) are given in italic}
\label{table:cashflows}
\end{table}

In this simplified setting the financing costs are not taken into account for both strategies. 

\section{Conclusions}\label{sec:conclusions}
In this study we investigated the implied correlation (IC) of the DAX index basket and introduced a hedging approach for the dispersion trading strategy using the IC forecast. We apply the dynamic semiparametric factor model (DSFM) to the IC dataset from January 2010 to August 2012, recover four basis functions and three time series of factors and use them to forecast the IC.  The advanced dispersion strategy we employ using the IC forecast shows the smallest maximal losses, the highest maximal gains and the highest average payoff over the studied sample period in comparison to the alternative strategies. So, we conclude that our modeling approach can be of potential use in equity dispersion trading.

The choice of DSFM as a model for the IC surface (ICS) dynamics is motivated by the degenerated dataset design, which has to be modeled nonparametrically. On the other hand we were driven by the necessity to reduce the dimensionality of the problem and facilitate the forecasting. DSFM satisfies both requirements. It captures the form of the ICS by its nonparametric part well, and allows a simple parametric model for dynamics to be used. At the later modeling stage we fit the three-dimensional VAR(2) model, which is a good choice to carry out the forecasting exercise. In addition, we found that it is possible to separate the influence of each recovered basis function on the ICS shape. The functions allow their interpretation as level, moneyness and maturity effects. The strength of these effects is defined by the time series of corresponding factors, which can be characterized as drivers of the correlation risk. An interesting task would be to study the presence, size and magnitude of the correlation risk premia, captured by these factors. We consider this findings to be important topics for further research.

\begin{acknowledgement}
The authors gratefully acknowledge financial support from the Deutsche Forschungsgemeinschaft through CRC 649 ``Economic Risk''.

Thanks go to L. Udvarhelyi for editorial assistance.
\end{acknowledgement}

\bibliographystyle{dcu}
\bibliography{bibliography}
\appendix
\newpage
\section{Appendix}

\subsection{Realized versus model free implied volatility}\label{append:crp}

\begin{table}[h!]
\begin{center}
\begin{tabular}{@{}m{5cm} *{3}{m{2cm}}@{}}
					&$\tau=0.25$	&$\tau=0.5$	&$\tau=1$	\\
\hline
BASF 				&	0.00000	&	0.00000	&	0.15298*\\
Commerzbank 			&	0.00000	&	0.00005	&	0.66073*\\
Continental AG 			&	0.99998*	&	0.99998*	&	0.99998*\\
Deutsche Bank 		&	0.00001	&	0.00000	&	0.06985*\\
Deutsche B\"orse 		&	0.95674*	&	0.95064*	&	0.96357*\\
Fresenius Medical Care 	&	0.45640*	&	0.27716*	&	0.81540*\\
Henkel 				&	0.90404*	&	0.99997*	&	0.99680*\\
K+S 					&	0.00000	&	0.00000	&	0.99981*\\
Lanxess 				&	0.27625*	&	0.05776*	&	0.99989*\\
Linde 				&	0.76162*	&	0.87214*	&	1.00000*\\
RWE 				&	0.40725*	&	0.21673*	&	0.05305*\\
ThyssenKrupp 			&	0.01733	&	0.00272	&	0.69073*\\
Volkswagen			&	0.99998*	&	0.99998*	&	0.99998*\\
\end{tabular}
\caption{The results of $t$-test for the equality of sample averages with $H_0: RV=MFIV$ against the alternative $RV<MFIV$ of DAX constituents for which the $H_0$ is not rejected at 5\% significance level at least for one $\tau$. * denotes test results, which cannot reject $H_0$. The test was performed for the volatilities of the DAX index and its 30 constituent stocks computed over the time period 20100104 - 20121228 for 3 different maturities/estimation windows: $\tau=0.25,0.5,1$. The test results are presented for the subsample of 13 DAX constituents, for which  $H_0$ cannot be rejected at least for one $\tau$.}\label{table:t_test_DAX}
\end{center}
\end{table}

\newpage
\subsection{Variance risk premium}\label{append:vrp}
\begin{table}[h!]
\begin{center}
\small
\begin{tabular}{@{}m{2cm}|*{3}{m{.75cm}}|*{3}{m{.75cm}}|*{3}{m{.75cm}}@{}}
 			&	\multicolumn{3}{c}{$\tau=0.25$}	&	\multicolumn{3}{c}{$\tau=0.5$}&	\multicolumn{3}{c}{$\tau=1$}\\
 			&	$\sigma$	&	$\tilde{\sigma}$	&	$\sigma-\tilde{\sigma}$			&	$\sigma$	&	$\tilde{\sigma}$	&	$\sigma-\tilde{\sigma}$			&	 $\sigma$	&	$\tilde{\sigma}$	&	$\sigma-\tilde{\sigma}$\\
\hline
Adidas 				&	27.56	&	30.06	&	-2.50	&	28.35	&	31.07	&	-2.71	&	29.72	&	31.45	&	-1.73\\
Allianz 				&	29.38	&	31.93	&	-2.55	&	30.44	&	33.03	&	-2.58	&	32.72	&	33.94	&	-1.22\\
BASF 				&	28.95	&	31.08	&	-2.13	&	29.73	&	31.62	&	-1.89	&	31.32	&	31.58	&	-0.26\\
Bayer 				&	27.39	&	30.74	&	-3.35	&	27.93	&	31.23	&	-3.31	&	28.85	&	31.17	&	-2.31\\
Beiersdorf 			&	33.90	&	37.09	&	-3.20	&	34.53	&	37.96	&	-3.43	&	35.95	&	37.94	&	-1.99\\
BMW 				&	19.57	&	23.95	&	-4.38	&	19.79	&	24.17	&	-4.38	&	20.26	&	23.90	&	-3.64\\
Commerzbank 			&	46.47	&	52.77	&	-6.30	&	47.40	&	52.54	&	-5.14	&	51.02	&	51.70	&	-0.68\\
Continental AG 			&	41.45	&	39.82	&	~1.63*&	43.84	&	40.55	&	~3.29*&	48.28	&	41.60	&	~6.69*\\
Daimler 				&	34.18	&	37.55	&	-3.36	&	34.93	&	38.45	&	-3.52	&	36.93	&	38.85	&	-1.93\\
Deutsche Bank 		&	39.26	&	43.56	&	-4.30	&	39.76	&	43.93	&	-4.18	&	42.52	&	43.51	&	-1.00\\
Deutsche B\"orse 		&	30.03	&	30.40	&	-0.37	&	31.09	&	31.10	&	~0.00*&	32.90	&	31.48	&	~1.43*\\
Deutsche Post 			&	31.61	&	33.63	&	-2.02	&	32.05	&	34.18	&	-2.13	&	33.13	&	34.67	&	-1.54\\
Deutsche Telekom 		&	25.50	&	27.54	&	-2.04	&	26.20	&	28.26	&	-2.06	&	27.66	&	29.12	&	-1.46\\
E.ON 				&	23.62	&	26.26	&	-2.64	&	24.11	&	26.53	&	-2.41	&	24.67	&	27.48	&	-2.81\\
Fresenius Medical Care 	&	28.12	&	29.21	&	-1.09	&	28.67	&	29.68	&	-1.01	&	30.19	&	30.04	&	~0.16*\\
Fresenius SE 			&	19.01	&	22.44	&	-3.43	&	19.21	&	23.17	&	-3.96	&	19.89	&	23.62	&	-3.73\\
HeidelbergCement 		&	21.16	&	23.56	&	-2.40	&	21.61	&	23.69	&	-2.08	&	22.77	&	24.08	&	-1.31\\
Henkel 				&	39.04	&	39.00	&	~0.05*&	40.96	&	39.49	&	~1.46*&	45.03	&	39.99	&	~5.04*\\
Infineon 	&	23.15	&	26.08	&	-2.94	&	23.57	&	26.70	&	-3.13	&	24.30	&	27.05	&	-2.75\\
K+S 					&	40.78	&	43.75	&	-2.98	&	41.75	&	45.03	&	-3.27	&	45.88	&	45.08	&	~0.80*\\
Lanxess 				&	29.99	&	30.84	&	-0.85	&	30.86	&	31.78	&	-0.92	&	33.51	&	32.52	&	~1.00*\\
Linde 				&	39.90	&	39.92	&	-0.02	&	41.02	&	40.55	&	~0.47*&	42.83	&	41.07	&	~1.76*\\
Lufthansa 				&	22.18	&	25.18	&	-2.99	&	22.72	&	26.32	&	-3.61	&	23.47	&	26.77	&	-3.30\\
Merck 				&	23.83	&	26.06	&	-2.23	&	24.43	&	26.40	&	-1.97	&	25.51	&	26.39	&	-0.89\\
Munich Re 			&	23.35	&	26.54	&	-3.19	&	23.88	&	27.91	&	-4.03	&	25.28	&	29.10	&	-3.82\\
RWE 				&	27.43	&	29.15	&	-1.72	&	27.93	&	29.71	&	-1.78	&	28.75	&	30.38	&	-1.63\\
SAP 					&	21.45	&	23.67	&	-2.22	&	21.73	&	25.14	&	-3.42	&	22.00	&	26.82	&	-4.82\\
Siemens 				&	25.94	&	28.53	&	-2.59	&	26.82	&	29.78	&	-2.95	&	28.63	&	30.35	&	-1.72\\
ThyssenKrupp 			&	36.32	&	38.37	&	-2.04	&	36.78	&	38.71	&	-1.93	&	38.80	&	38.90	&	-0.10\\
Volkswagen			&	37.91	&	36.31	&	~1.61*&	39.48	&	36.90	&	~2.59*&	41.94	&	36.91	&	~5.03*\\
\textbf{DAX Index}		&	21.72	&	25.38	&	-3.66	&	22.30	&	26.67	&	-4.37	&	23.40	&	27.81	&	-4.41\\
\end{tabular}
\caption{Mean of the $\sqrt{RV}$ ($\sigma$) and $\sqrt{MFIV}$ ($\tilde{\sigma}$) and their difference $\sqrt{RV}-\sqrt{MFIV}$ ($\sigma-\tilde{\sigma}$), for DAX index and its 30 constituent stocks computed over the time period 20100104 - 20121228 for 3 different maturities/estimation windows: $\tau=0.25,0.5,1$). $\sigma-\tilde{\sigma}\geq0$ are marked with *.\label{table:mean_rv_mfiv_DAX}}
\end{center}
\end{table}


\newpage
\subsection{Dependence of index volatility and equicorrelation}\label{apsec:breakpoints}
\begin{figure}[h!]
\begin{center}
\includegraphics[width=0.9\textwidth]{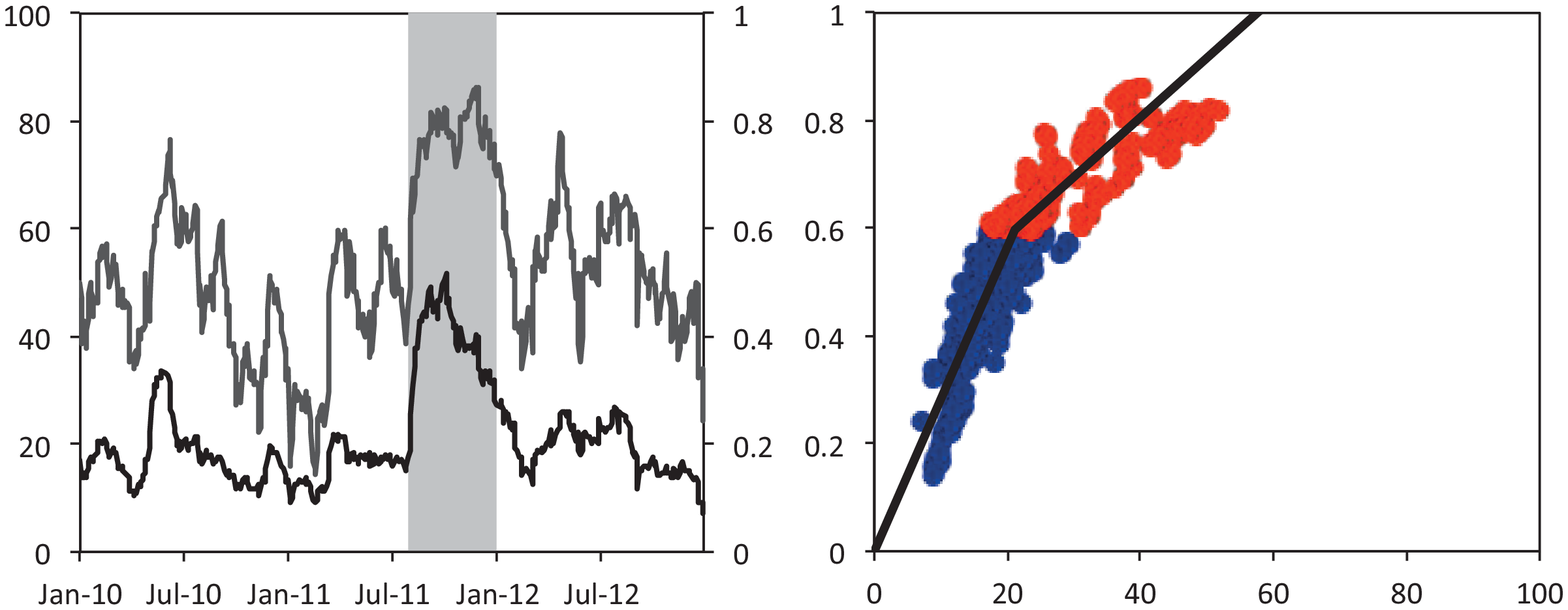}
\includegraphics[width=0.9\textwidth]{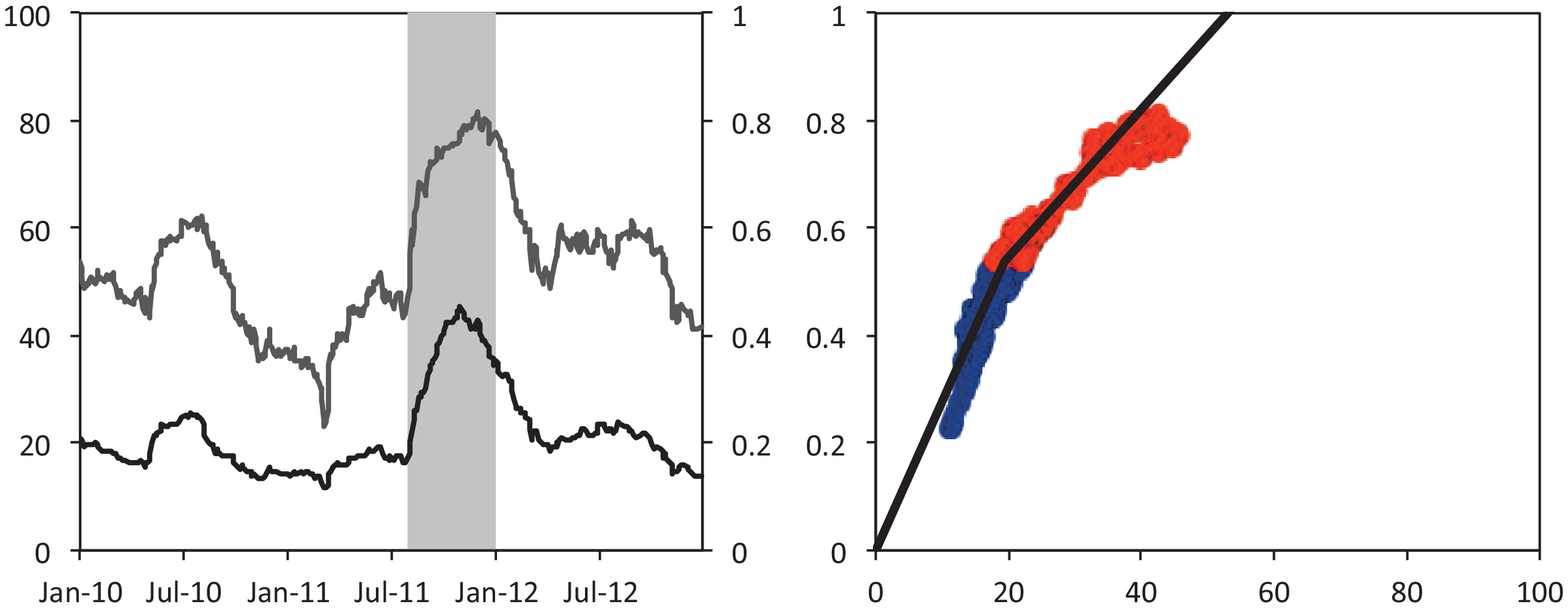}
\includegraphics[width=0.9\textwidth]{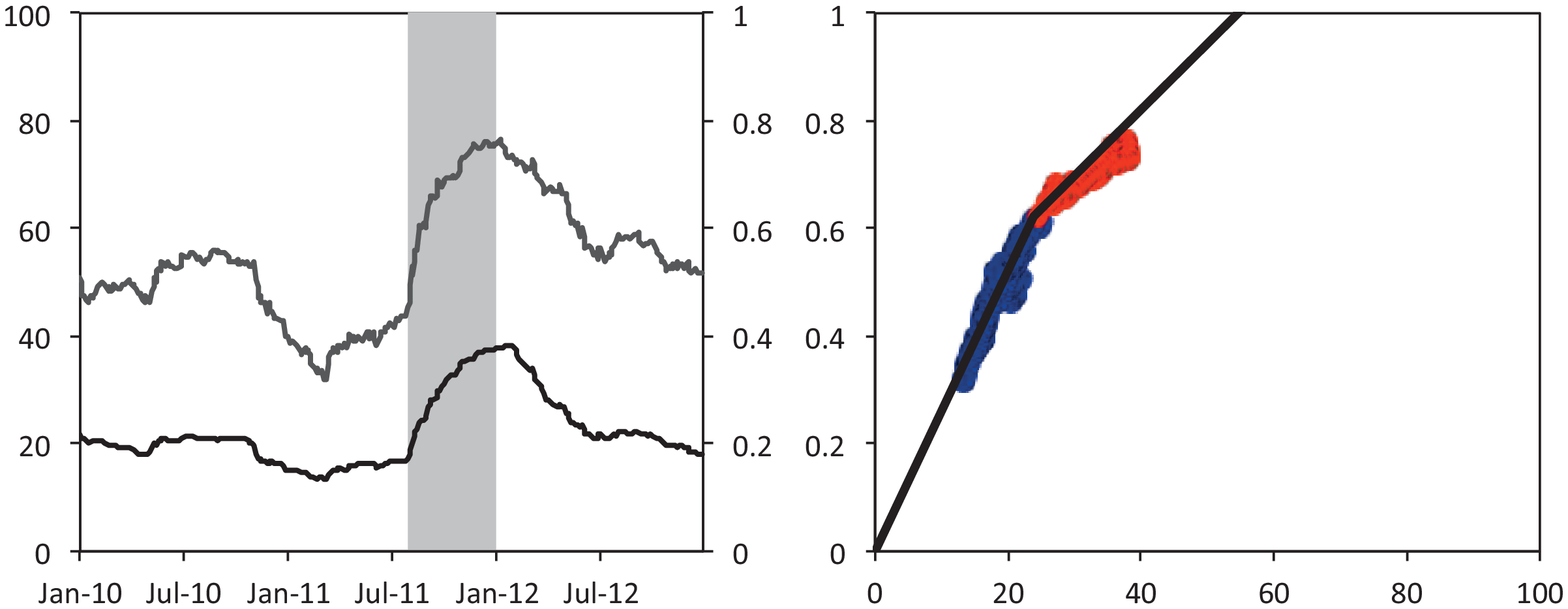}
\caption{DAX $\sigma_{B,t,\tau}$ (solid line) vs $\rho_{t,\tau}$ (dashed line), scatter plot $\sigma_{B,t,\tau}$ vs $\rho_{t,\tau}$, for $t,\tau$ from 20100104 till 20121228,  estimated with (\ref{eq:rv}) and (\ref{eq:eqcorr}) over 1 month ($\tau=0.083$), 3 months ($\tau=0.25$) and 6 months ($\tau=0.5$) window. Shaded area: Aug 2011 market fall. Switch point for two regression line is defined as described in Section \ref{sec:data}. \label{fig:example_RC_RV_6m}}
\end{center}
\end{figure}
\newpage
\begin{figure}[h!]
\begin{center}
\includegraphics[width=0.9\textwidth]{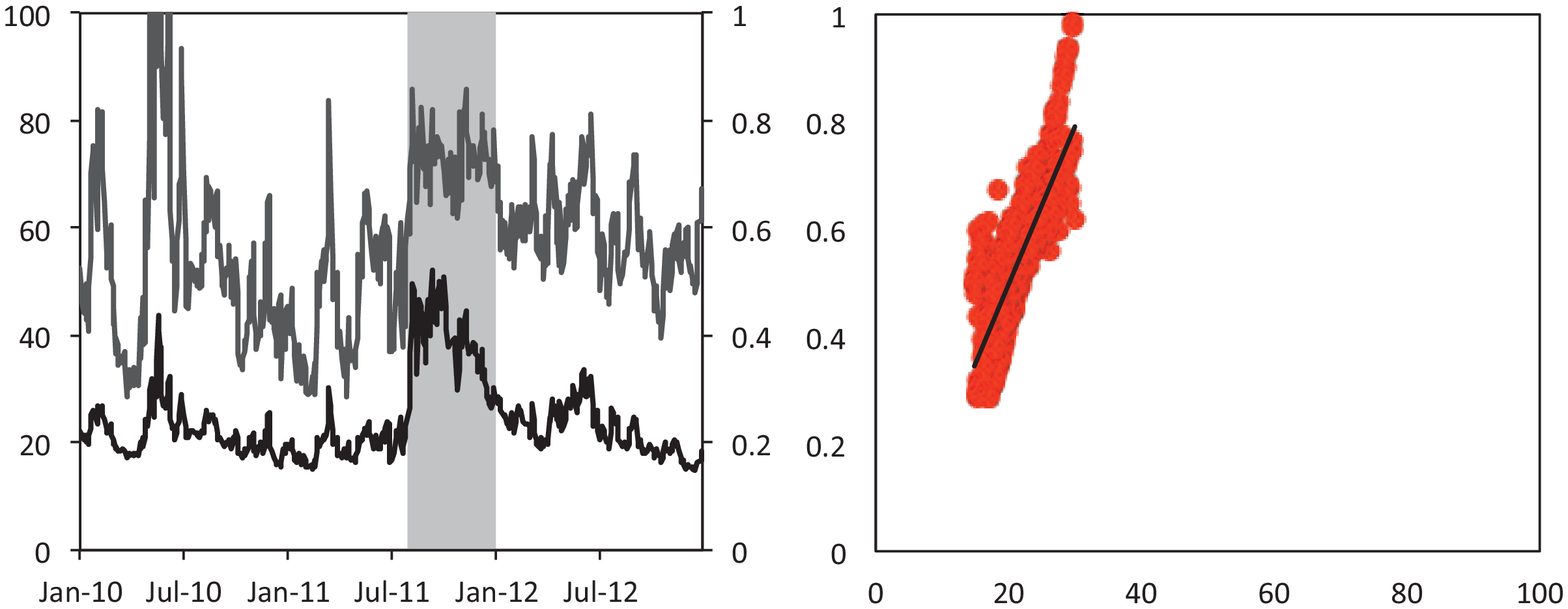}
\includegraphics[width=0.9\textwidth]{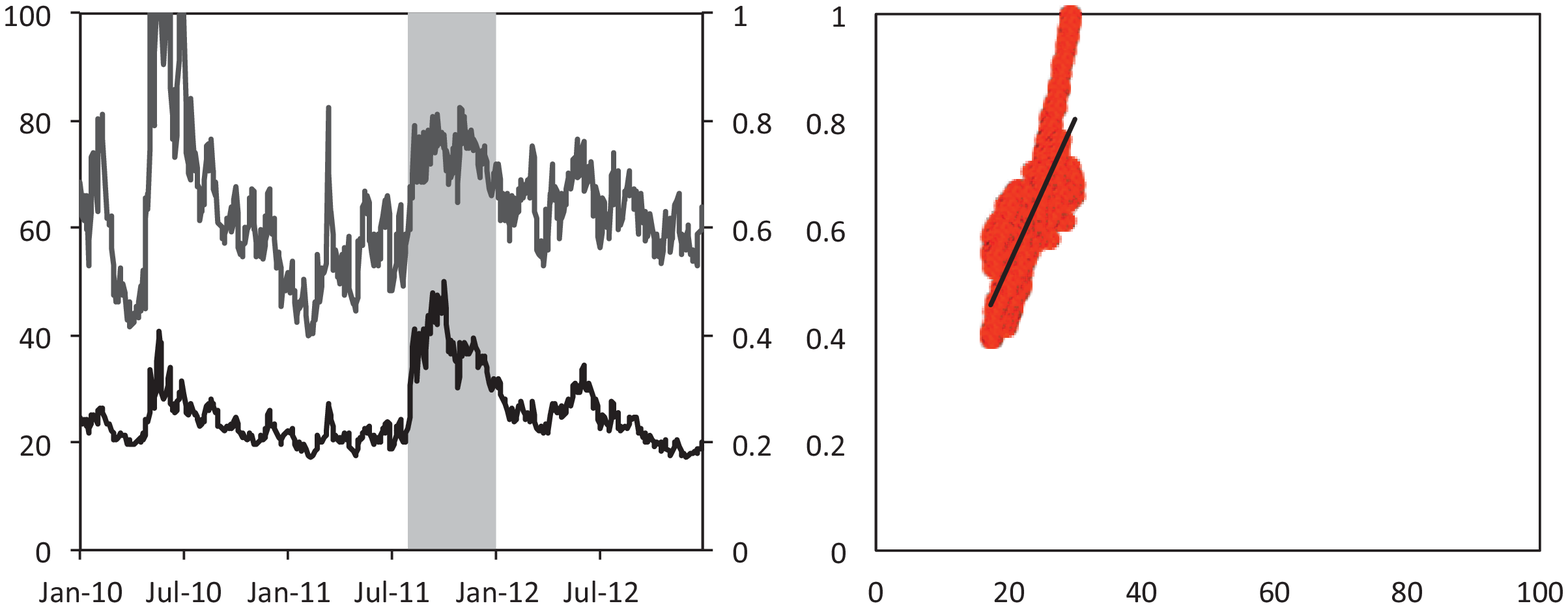}
\includegraphics[width=0.9\textwidth]{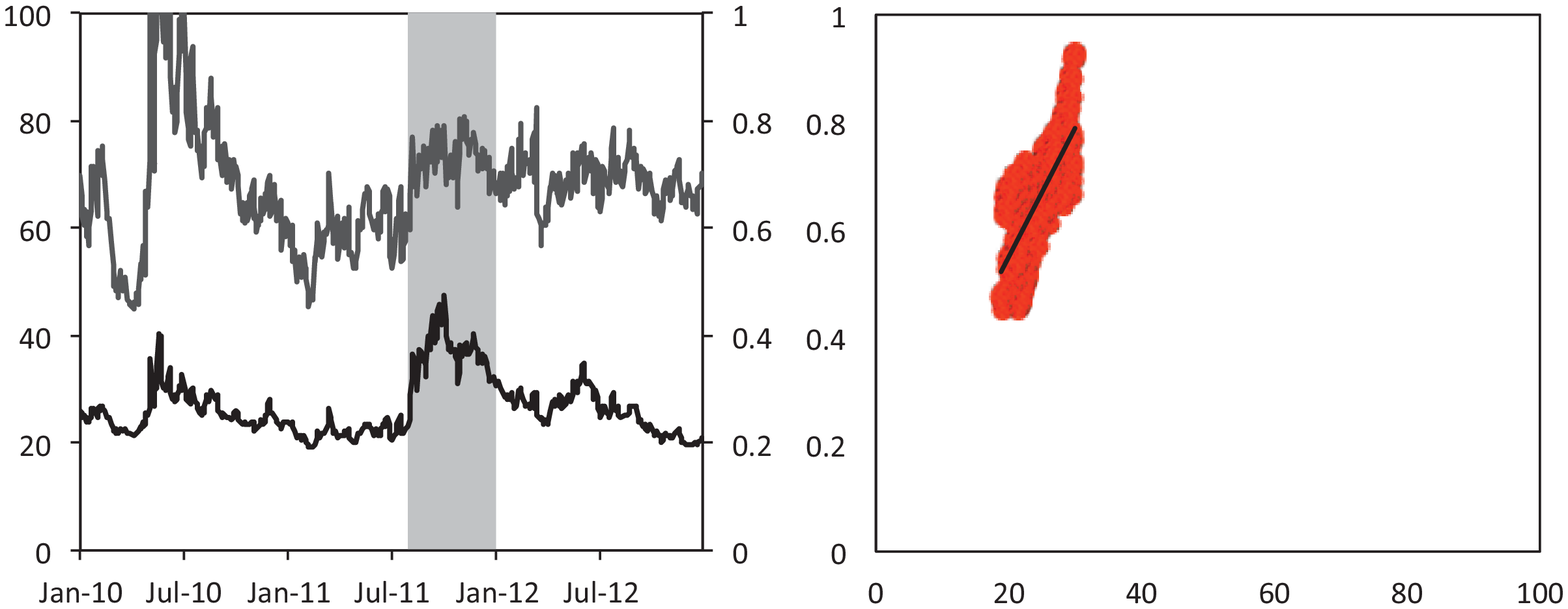}

\caption{DAX $\widehat{\sigma}_{t,B}(1,\tau)$ (solid line) vs $\widehat{\rho}_{t}(1,\tau)$ (dashed line), scatter plot $\widehat{\sigma}_{t,B}(1,\tau)$ and $\widehat{\rho}_{t}(1,\tau)$, for $t,\tau$ from 20100104 till 20121228,  estimated from IVs with (\ref{eq:ics}) for option with 1 month ($\tau=0.083$), 3 months ($\tau=0.25$) and 6 months ($\tau=0.5$) maturity. Shaded area: Aug 2011 market fall.\label{fig:example_MFIC_MFIV_6m}}
\end{center}
\end{figure}

\newpage
\subsection{Switch point selection for correlation regimes}\label{append:switchpoint}

\begin{table}[h!]
\begin{center}
\begin{tabular}{c|*{4}{c}}
$\tau$	&	$\sigma_{B,t+\tau}$	&	$\rho_{t+\tau}$	&	Slope 1	&	Slope 2\\
\hline

0.083		&	20.24			&	0.5917		&	0.0361	&	0.0085\\
0.25		&	20.34 		&	0.5728		&	0.0336	&	0.0093\\
0.5			&	22.42			&	0.6008		&	0.0286	&	0.0094\\
\hline 
Average	&	21.00			&	0.5884		&	0.0328	&	0.0091\\
\end{tabular}
\caption{Segmented linear regression of $\rho_{t+\tau}$ on $\sigma_{B,t+\tau}$ with one break point, $\tau=0083, 0.25, 0.5$ for $t+\tau$, from 20100104 till 20120801}
\label{table:CHANGE_POINT}
\end{center}
\end{table}
The dependence of $\rho$ and $\sigma_{B}$ observed in RV and RC is not pronounced in case of ATM IV and IC (Appendix \ref{apsec:breakpoints}).
Therefore we propose a market regime correction scheme for the IC dataset. 
The idea is to find a breakpoint between two segments of a piecewise linear regression of $\rho_{t+\tau}$ on $\sigma_{B,t+\tau}$. Using the procedure described in \cite{Muggeo2003} we fit a segmented linear regression with one break point. Based on results summarized in Table \ref{table:CHANGE_POINT} we make a following state dependent correction: if $\widehat{\sigma}_{B,t}(1,\tau)>21$ (high volatility regime), then $\widehat{\rho}_{t}(\kappa,\tau)=0.0091\widehat{\sigma}_{B,t}(\kappa,\tau)$

\subsection{Smoothing Parameters Selection}\label{apsec:badnwidth}

For both (\ref{eq:1est_step}) and (\ref{eq:2est_step}) kernel bandwidths ${h_\mu=(h_{\mu,1},h_{\mu,2})^{\top}}$ and ${h_\phi=(h_{\phi,1},h_{\phi,2})^{\top}}$ are to be selected.  As suggested in \cite{Haerdle2004} we use  the penalizing function approach to select optimal $h^{opt}_{\mu}$, minimizing mean integrated squared error (MISE):

\begin{equation}\label{eq:aic}
\frac{1}{T}
\sum_{t=1}^T \frac{1}{J_t} \sum_{j=1}^{J_t}
\left \{Y_{t,j} - \sum_{l=1}^L \widehat Z_{t,l} \widehat m_l(X_{t,j}) \right \}^2
w_{h^*,t}(X_{t,j})
\Xi_{AIC}
\left \{\frac{W_{h^*,t,j}(X_{t,j})}{TJ_t}  \right \},
\end{equation}

with the Akaike (1970) Information Criterion (AIC) as penalizing function $\Xi_{AIC}(q)=\exp(2q)$ and $W_{h^*,t,j}(X_{t,j})$ defined by

\begin{equation}\label{eq:W}
W_{h^*,t,j}(X_{t,j})=\frac{\mathcal{K}_{h}(0)}{J_t^{-1}\sum_{k=1}^{J_t} \mathcal{K}_{h}\left(X_{t,k}-X_{t,j}\right)},
\end{equation}
for every $X_{t,j}$, $1\leq t \leq T$, $1\leq j \leq J_t$.

Since the distribution of the observations is very uneven, we are using the weighted version of the criterion with weights $w_{h^*,t}(\bar{u})\stackrel{\operatorname{def}}{=}p_{h^*,t}^{-1}(\bar{u})$, where $p_{h^*,t}(\bar{u})$ is the average design density. For every $X_{t,j}$, $1\leq t \leq T$, $1\leq j \leq J_t$ it is defined by:

\begin{equation}\label{eq:add}
p_{h^*,t}(X_{t,j})=J_t^{-1}\sum_{k=1}^{J_t} \mathcal{K}_{h}\left(X_{t,k}-X_{t,j}\right),
\end{equation}

The bandwidth $h^{opt}_{\mu AIC}=(h_{\mu 1},h_{\mu 2})^{\top}$ corresponding to the minimal criterion \ref{eq:aic} is taken as optimal. The bandwidth $h^*$ of the weighting function is constant and does not depend of choice of ${h_\mu}$.

\subsection{Testing Factor Time Series for Stationarity}\label{apsec:nlags}

\begin{table}[h!]
\begin{center}
\begin{tabular}{l *{3}{r @{.} l @{ } l}}
&\multicolumn{3}{c}{$\widehat Z_{t,1}$}	&	\multicolumn{3}{c}{$\widehat Z_{t,2}$}	&	\multicolumn{3}{c}{$\widehat Z_{t,3}$}\\
\hline
20090803 - 20100730 (the 1st year)	&-2&991&(1)			&	-6&982&(1)		&	-5&710&(3) \\
20100802 - 20110729 (the 2nd year)	&-1&666*&(3)		&	-3&090&(2)		&	-4&480&(1) \\
20110802 - 20120801 (the 3rd year)	&-3&511&(2)			&	-3&796&(3)		&	-3&480&(2)\\
20090803 - 20120801 (the entire sample)	&-4&025&(1)			&	-6&912&(3)		&	-8&979&(1)\\
\end{tabular}
\end{center}
\caption{Augmented Dickey-Fuller (ADF) test on $\widehat Z_{t,1},\widehat Z_{t,2},\widehat Z_{t,3}$. Number of lags included in the ADF regression (in brackets) is chosen by starting with 3 lags and subsequently deleting lag terms, until the last one is significant at 5\% level. Test statistics that does not reject the hypothesis of a unit root at 5\% level are denoted by *.\label{tab:ADF_DAX}}
\end{table}

\newpage
\subsection{Determining the Number of Lags for VAR Model}\label{apsec:nags}

\begin{table}[h!]
\begin{center}
\begin{tabular}{*{2}{l}  *{3}{ r@{.} l}}
&&\multicolumn{2}{c}{AIC}&\multicolumn{2}{c}{HQIC}&\multicolumn{2}{c}{SBIC}\\
\hline
20090803 - 20100730						& 1	&1&923	&	2&061	&	2&162\\
(the 1st year) 								& 2	&1&839*	&	1&975*&	2&152*\\
										& 3	&1&856	&	2&052	&	2&304\\
										& 4	&1&882	&	2&060	&	2&389\\
\hline			
20100802 - 20110729 						& 1	&-2&868	&	-2&800	&	-2&699\\
(the 2nd year)								& 2	&-3&075*&	-2&932*	&	-2&755*\\
										& 3	&-3&068	&	-2&898	&	-2&645\\
										& 4	&-3&051	&	-2&854	&	-2&525\\
\hline			
20110802 - 20120801 						& 1	&-0&118	&	-0&051	&	0&048\\
(the 3rd year)			 					& 2	&-0&355	&	-0&238*	&	-0&064*\\
										& 3	& 0&361*&	-0&193	&	0&055\\
										& 4	&-0&360	&	-0&144	&	0&179\\
\hline
20090803 - 20120801 						& 1	&0&745	&	0&773	&	0&818\\
(the entire sample)							& 2	&0&384*	&	0&461*	&	0&539*\\
										& 3	&0&397	&	0&467	&	0&579\\
										& 4	&0&412	&	0&475	&	0&621\\
\end{tabular}
\end{center}
\caption{Akaike's information criterion (AIC), Schwarz's Bayesian information criterion (SBIC), and the Hannan and Quinn information criterion (HQIC) for defining the optimal lag order $p$ of a VAR model for DAX and S\&P100 ICS  factors $\widehat Z_{t,1},\widehat Z_{t,2},\widehat Z_{t,3}$. * appearing next to the test statistics indicates the optimal lag at 5\% significance level. \label{tab:lag_order}}
\end{table}

\newpage
\subsection{Parameters of VAR model for  DAX ICS factors}\label{apsec:DAX_VAREST}
\begin{table}[h!]
\begin{center}
\begin{tabular}{l *{3}{ r@{.} l}|*{3}{ r@{.} l}}
&\multicolumn{6}{c}{20090803 - 20120801}&\multicolumn{6}{c}{20090803 - 20100730}\\
&\multicolumn{6}{c}{(the entire sample)}&\multicolumn{6}{c}{(the 1st year)}\\
\hline
&\multicolumn{2}{c}{$Z_{1,t}$}&\multicolumn{2}{c}{$Z_{2,t}$}&\multicolumn{2}{c}{$Z_{3,t}$}&\multicolumn{2}{c}{$Z_{1,t}$}	&\multicolumn{2}{c}{$Z_{2,t}$}&\multicolumn{2}{c}{$Z_{3,t}$}\\
\hline
$Z_{1,t-1}$	&	0&645*	&	-0&012	&	-0&019	&	0&630*	&	-0&032	&	0&029\\
$Z_{1,t-2}$	&	0&310*	&	0&008	&	0&029	&	0&276*	&	0&013	&	-0&060*\\
$Z_{2,t-1}$	&	-0&104*	&	0&259*	&	0&156	&	-0&036	&	0&047	&	0&036\\
$Z_{2,t-2}$	&	0&057*	&	0&406*	&	-0&014	&	-0&039	&	0&339	&	-0&104*\\
$Z_{3,t-1}$	&	-0&07	&	0&140*	&	0&471*	&	-0&091	&	-0&494*	&	0&525*\\
$Z_{3,t-2}$	&	0&149*	&	0&118*	&	0&251*	&	0&046	&	0&181	&	-0&208*\\
$c$			&	0&004	&	0&006	&	-0&003	&	-0&004	&	-0&001	&	0&001\\
\hline
&\multicolumn{6}{c}{20100802 - 20110729}&\multicolumn{6}{c}{20110802 - 20120801}\\
&\multicolumn{6}{c}{(the 2nd year)}&\multicolumn{6}{c}{(the 3rd year)}\\
\hline
&\multicolumn{2}{c}{$Z_{1,t}$}&\multicolumn{2}{c}{$Z_{2,t}$}&\multicolumn{2}{c}{$Z_{3,t}$}&\multicolumn{2}{c}{$Z_{1,t}$}&\multicolumn{2}{c}{$Z_{2,t}$}&\multicolumn{2}{c}{$Z_{3,t}$}\\
\hline
$Z_{1,t-1}$	&	0&809*	&	0&048	&	-0&202*	&	0&339*	&	0&191*	&	-0&001\\
$Z_{1,t-2}$	&	0&254*	&	-0&029	&	0&112*	&	0&351*	&	0&041	&	-0&036\\
$Z_{2,t-1}$	&	0&188*	&	0&687*	&	-0&223*	&	0&355*	&	0&264*	&	0&132*\\
$Z_{2,t-2}$	&	0&091	&	0&262*	&	0&018	&	0&084	&	0&302*	&	-0&045\\
$Z_{3,t-1}$	&	0&453*	&	0&051	&	0&162*	&	-0&197	&	-0&008	&	0&623*\\
$Z_{3,t-2}$	&	0&118	&	0&118	&	0&275*	&	0&044	&	0&240*	&	0&204*\\
$c$			&	-0&004	&	0&001	&	-0&002	&	0&003	&	-0&001	&	-0&002\\
\end{tabular}
\end{center}
\caption{The estimated parameters for VAR(2) model for DAX ICS factors. * marks estimates which are not significant at 5\% level.\label{tab:DAX_VAREST_VAR2}}
\end{table}

\end{document}